\documentclass[12pt,preprint]{aastex}

\def\ltsima{$\; \buildrel < \over \sim \;$}
\def\simlt{\lower.5ex\hbox{\ltsima}}
\def\gtsima{$\; \buildrel > \over \sim \;$}
\def\simgt{\lower.5ex\hbox{\gtsima}}

\shorttitle{SFH of the LMC}
\shortauthors{Smecker-Hane et al.}

\begin{document}

\title{The Star Formation History of the Large Magellanic Cloud 
\altaffilmark{1, 6}}

\author{Tammy A. Smecker-Hane\altaffilmark{2},
Andrew A. Cole\altaffilmark{3},
John S. Gallagher, III\altaffilmark{4},
and Peter B. Stetson\altaffilmark{5}}

\altaffiltext{1}{Based on observations made with the NASA/ESA
{\it Hubble Space Telescope}, obtained at the Space Telescope
Science Institute, which is operated by the Association of
Universities for Research in Astronomy, Inc., under NASA contract
5-26555.}
\altaffiltext{2}
{Department of Physics and Astronomy, University of California,
Irvine, 4129 Frederick Reines Hall, Irvine, CA 92697--4575;
{\it tsmecker@uci.edu}}
\altaffiltext{3}{Department of Astronomy, 532A Lederle Grad Research Tower,
University of Massachusetts, Amherst, MA 01003;
{\it cole@condor.astro.umass.edu}}
\altaffiltext{4}
{Department of Astronomy, University of Wisconsin--Madison,
5534 Sterling Hall, 475 North Charter Street, Madison, WI 53706--1582;
{\it jsg@astro.wisc.edu}}
\altaffiltext{5}
{Dominion Astrophysical Observatory, Herzberg Institute of
Astrophysics, 5071 West Saanich Road, Victoria, BC V8X 4M6, Canada;
{\it peter.stetson@hia.nrc.ca}}
\altaffiltext{6}
{Accepted for publication in the Astrophysical Journal}

\begin{abstract}
Using WFPC2 aboard the {\it Hubble Space Telescope}, we have
created deep color-magnitude diagrams in the V and I passbands
for approximately $10^5$ stars in a field at the center of the LMC bar 
and another in the disk.  The main--sequence luminosity functions (LFs) 
from 19 $\leq$ V $\leq$ 23.5, the red clump and horizontal branch
morphologies, and the differential Hess diagram of the two fields
all strongly imply that the disk and bar have significantly different
star-formation histories (SFHs). The disk's SFH has been relatively
smooth and continuous over the last $\sim 15$ Gyr while the bar's 
SFH was dominated by star formation episodes at intermediate ages.
Comparison of the LF against predictions based on Padova theoretical 
stellar evolution models and an assumed age-metallicity relationship 
allows us to identify the dominant stellar populations in the bar with 
episodes of star formation that occurred from 4 to 6 and 1 to 2 Gyr ago.  
These events accounted, respectively, for $\sim$ 25\% and $\sim$ 15\% 
of its stellar mass.  The disk field may share a mild enhancement
in SF for the younger episode, and thus we identify the 4 to 6 Gyr
episode with the formation of the LMC bar.
\end{abstract}



\section{Introduction}

The Large Magellanic Cloud (d $\approx$ 50 kpc) 
is the nearest galaxy to our own that has significant mass
($\sim$10$^{10}$ M$_{\sun}$). 
Its small distance allows individual stars to be resolved and makes it
an ideal target for studying its global SFH
and chemical evolution.  The detailed SFH of the LMC is tractable now 
because of the high angular resolution imaging possible with the 
{\it Hubble Space Telescope} (e.g., Gallagher et al.~1999 and 
references therein), 
and the potential for obtaining chemical abundances 
for many individual stars with new multi-object spectrographs on 
ground-based 4-meter telescopes.
Even so, the global history of the LMC is not yet known with significant
precision or age resolution.

The Wide--Field/Planetary Camera 2 (WFPC2) aboard HST allows the photometric
study of main--sequence stars with masses as low as 0.6 M$_{\sun}$,
i.e., stars which have not evolved significantly during the lifetime
of the LMC \citep{gal96}.   However, the ability to draw 
conclusions about the global history of the LMC is hampered by the
extremely small size of the WFPC2 field of view, which results
in a large uncertainty due to small number statistics
in analyses of color-magnitude diagrams (CMDs).  To overcome this 
limitation, we have obtained deep images for numerous WFPC2 pointings in
the LMC bar and disk and produced CMDs with $\approx 10^5$ stars
per field.  In order to overcome the degeneracy of age and metallicity
in CMDs, we are directly measuring the metallicity distribution
by determining chemical abundances with 0.2 dex precision
for hundreds of red giant stars in these fields using the strength of
the calcium infrared triplet lines (Cole, Smecker-Hane \& Gallagher 2000 
[CSG00], Smecker-Hane et al.~2001). By combining the 
metallicity distributions with CMDs, we will obtain the clearest picture 
yet of the evolution of the LMC.

Up to now, the SFH of the LMC's bar, as distinct from its disk, 
has been poorly constrained.
In this {\it Letter}, we report the first results of our program ---  a
direct comparison of the CMDs of the bar and disk fields, which
shows that the LMC bar has experienced a significantly
different SFH than the disk. We also model the main--sequence luminosity
functions to derive the SFHs.  A more detailed analysis 
of the CMDs incorporating our spectroscopic results
will be reported in future papers.

\section{Data: Observations, Reductions, Photometry}

Our data were obtained in 1997 October--November  
under HST Guest Observer Proposal \#7382 (PI = Smecker-Hane).
The bar field adopted here
consists of 2 out of a total of 
4 WFPC2 pointings near the open cluster HS 275
($\alpha$ = 5$^h$24$^m$, $\delta$ = $-$69$\arcdeg$46$\arcmin$
[J2000.0]), near the optical center of
the bar.  The disk field, referred to as ``Disk 1'' (CSG00),
lies 1$\fdg$7 southwest of the center of the LMC along the
direction of the bar's minor axis, near the open cluster 
SL 336, ($\alpha$ = 5$^h$14$^m$, $\delta$ = 
$-$71$\arcdeg$13$\arcmin$, [J2000.0]). The Disk 1 field is well outside
the bar, and hence the CMD produced here samples a pure disk population.
In Disk 1, 10 WFPC2 pointings were mosaiced in order
to sample about as many stars as in the bar field.  Each pointing 
consisted of 4$\times$500 sec integrations in the F555W filter
and 2$\times$300 sec and 2$\times$700 sec in F814W.  Exposures were
dithered in non-integer pixel amounts to mitigate the undersampling 
of the point-spread function.
The data were reduced in the standard HST pipeline.
Profile-fitting photometry was performed on the individual images 
using DAOPHOT II/ALLFRAME software
\citep{ste87,ste94}, and calibrated to Johnson V and Kron-Cousins I
magnitudes \citep{hol95}.  Further details of our reductions, 
photometry, and artificial star tests will be reported in 
a subsequent paper; here we report our most important early results.

\section{Color-Magnitude Diagrams}

\begin{figure}
\plotone{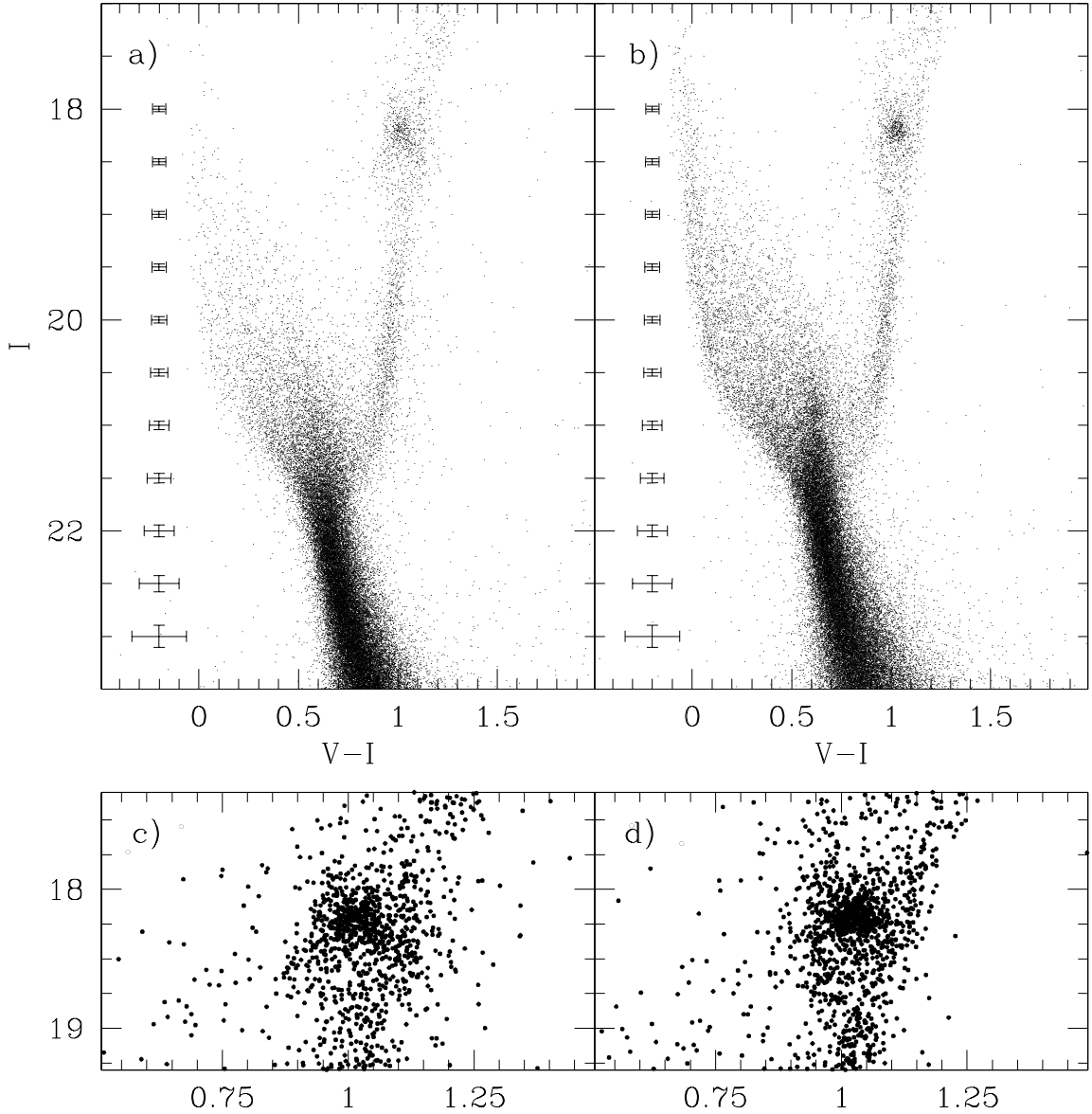}
\caption{WFPC2 CMDs:  {\it a)} the Disk 1 field located 1$\fdg$7
from the center of the LMC;   {\it b)} the Bar field.
Representative errorbars are shown as a function of magnitude.
Panels {\it c)} and {\it d)} magnify the red
clump region to facilitate a detailed comparison.
\label{cmds}}
\end{figure}

The CMDs for the Disk 1 and Bar
fields are shown in Figure~\ref{cmds}. They extend more than 2.5
magnitudes below the oldest main--sequence turnoffs (MSTOs) in the LMC.
The CMDs are similar in their gross properties, showing a strong
upper main sequence of youthful stars, prominent
red clumps characteristic of intermediate-age populations, and 
corresponding structures in the MSTO/subgiant
region suggesting continuous star formation with resulting
chemical enrichment.  Photometric completeness in the bar
field drops below 50\% for I $\geq 24.0$, and
saturation occurs for V $\leq 18.0$.  Thus we focus our analysis 
on the intervening region, where our photometry is reliable and 
complete.

Although broadly similar, the two CMDs differ in a number of significant 
ways.  In the bar field, the main sequence brighter than I $= 20.5$ 
is much more concentrated towards the zero--age main sequence 
suggesting a larger 
fraction of young stars. Note that, because of saturation,  we can 
measure only MSTOs with ages $\simgt$ 200 Myr. A small fraction
of the stars in the bar field is younger than this, but such a
population is weaker in Disk 1.  The red horizontal branch
region of the Disk 1 CMD ($0.6 \simlt$ V$-$I $\simlt 0.9$, 
$18.5 \simlt$ I $\simlt 19$) is sparsely populated, but 
more richly than the corresponding region of the bar CMD.
This implies a deficiency of metal-poor, ancient (t $\geq$ 10 Gyr) stars
relative to younger stars in the bar, when compared to Disk 1.
The red clump morphologies of the
two fields differ strongly, as discussed below and shown in 
Figure~\ref{cmds}c--d.  Perhaps most significantly, 
the bar field
shows a large excess of stars at V$-$I $\approx$ 0.65 from
$20.5 \simlt $ I $\simlt 21.5$.

These differences in CMD morphology are {\it independent} of assumed distance, 
mean reddening and stellar evolution models.  However, their detailed
quantitative interpretation requires us to adopt values for 
these.  We have estimated a reddening of E(B$-$V) $= 0.03$ for
Disk 1 based on Str\"{o}mgren photometry (CSG00); by comparing the colors 
of the upper main sequences, we find the bar field has slightly higher 
reddening, E(B$-$V) $\approx$ 0.05.  The distance to the LMC is controversial
(see, e.g., Cole 2000 and references therein), but the assumption
of a fiducial (m$-$M)$_0$ = 18.5 is reasonable.
Our CMDs are not ideal for estimating the stellar metallicities in the fields.
However, the colors of the red clump and lower main sequence 
can be used to infer that Disk 1 and the bar have roughly similar
metallicities.  For reference, from spectra of 39 red giants in Disk 1, 
we find a mean metallicity of [Fe/H] $\approx$ $-$0.6 (CSG00).

\section{Bar and Disk: A Comparative Analysis}

\begin{figure}
\plotone{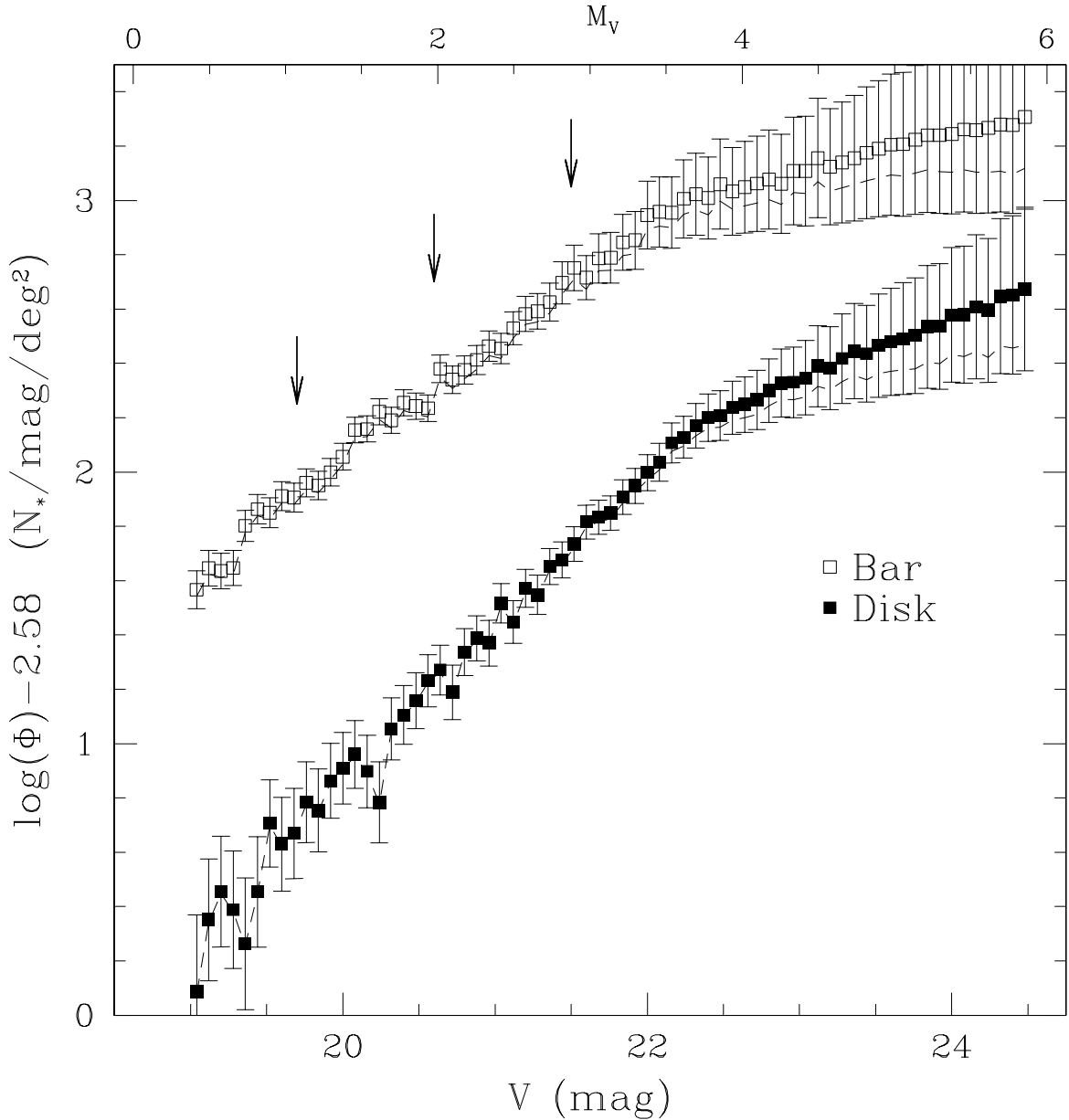}
\caption{ Main--sequence LFs of the LMC bar and Disk 1 fields. 
The raw LFs, uncorrected for incompleteness,
are shown as dashed lines.  Bin sizes are 0.05 mag.  The LFs are 
shown normalized to the same stellar surface density; the zeropoint 
of the ordinate is set by the area observed in the bar field, 
9.4 arcmin$^2$. The arrows denote spikes in the bar LF with associated
changes in LF slope that imply enhancements in the SFR.
\label{lffig}}
\end{figure}

We begin by 
examining the main--sequence luminosity functions (LFs) of 
the two fields.  These are presented in Figure~\ref{lffig};
note that we give the LF in units of number of star per magnitude
per square area to account for the differences in area surveyed
in each field.  The LFs have been corrected for 
incompleteness using artificial star tests;
the uncorrected LFs are shown as dashed lines. Error bars show the
1$\sigma$ errors derived from Poisson statistics.
We find that:
1) the bar and disk LFs both show a ``knee" at V $\approx$ 22.2,
which indicates the oldest MSTO of the LMC has an age $> 10$ Gyr; 
2) the bar LF is shallower than the Disk 1 LF, implying a larger
fraction of young and intermediate-age stars in the bar; 
3) the bar LF shows large variations compared to the relatively 
smooth Disk 1 LF, which translates to a more burst-dominated SFH 
for the bar.  The knees in the LFs occur at fainter magnitudes 
than had been derived using ground-based telescopes; this shows that even 
for relatively bright stars, the high resolution obtainable
from space is vital for accurate photometry.

The spikes in the bar LF at V $= 21.5$,
20.6, and 19.7, and the corresponding changes in the LF slope
at those magnitudes, indicate large temporal variations in the 
star-formation rate (SFR).
Smaller amplitude variations, of marginal statistical significance, 
are seen in the Disk 1 LF; note the almost complete lack
of a feature corresponding to the peak in the bar LF at V = 21.5.
This provides {\it model-independent} evidence that a major
episode of star formation, beginning several Gyr after the 
oldest LMC stars formed, occurred in the bar.

To derive the SFH from the LFs, we need to adopt three key 
ingredients: 1) a stellar initial mass function,
2) a set of stellar evolutionary models, and 3) an age--metallicity 
relationship. We assume the initial mass function
from \citet{kro93}. We adopt the newest stellar evolutionary 
models from the Padova group (Girardi et al.~2000). Using
equivalent evolutionary points, we interpolate
them to arbitrary age and metallicity. We began our analysis by 
adopting the age-metallicity relationship of \citet{pag98}, but
comparisons of the observed CMDs to model CMDs showed
that the observed CMDs clearly did not have as many metal-poor stars 
as predicted.  (The observed red giant branch was narrower and redder
than predicted, and the observed CMD lacked the extended blue HB 
predicted by the models.) We believe the lack of metal--poor stars in 
these fields is indeed real because we find very few stars with 
metallicities Z $\leq 0.001$ in our spectroscopic surveys of red 
giant stars in the Disk 1 field (CGS00), and in our subsequent 
work in the Disk 1 field and another disk field at similar 
galactocentric radius (Smecker-Hane et al.~2001).  Therefore, we 
have adopted the age--metallicity relationship from \citet{pag98} 
for ages $\leq 10$ Gyr and assumed a constant metallicity, Z $=0.003$, 
for ages $\geq  10$ Gyr. In future papers, we will derive both
the age--metallicity relationship and the SFH in a self-consistent
way by simultaneously modeling the CMDs and metallicity distribution 
functions.

\begin{figure}
\plotone{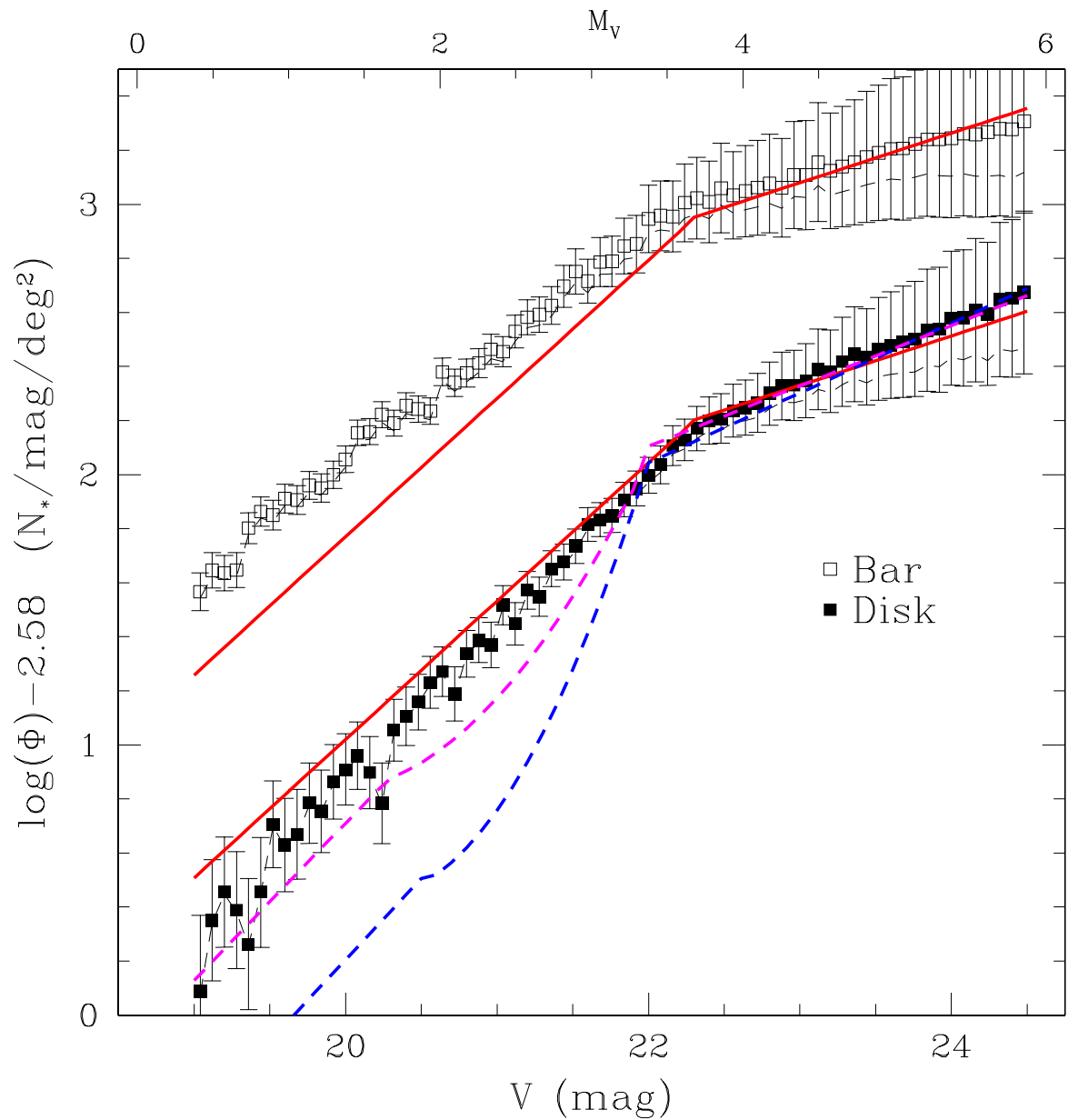}
\caption{ Main--sequence LFs of the bar and Disk 1 fields.
Model LFs for a constant SFR (solid, red lines), 
and exponentially declining SFR with $\tau = 8$ Gyr (dashed, magenta line) 
and $\tau = 4$ Gyr (dashed, blue line). See text for additional details.
\label{lfmod}}
\end{figure}

Figure \ref{lfmod} compares the observed and model LFs. 
We show three model LFs computed assuming a constant SFR with 
the onset of star formation, $t_0$, assumed to be 14.2 Gyr ago, and
exponentially declining rates, SFR $\propto e^{-(t-t_0)/\tau}$,
with $\tau = 8$ and 4 Gyr. For comparison with observed 
LFs, model LFs have been normalized to have the same number of stars
in the magnitude range $22 \leq V \leq 24$. A constant SFR gives 
$\log \Phi(V) \propto 0.51 V$.  The Disk 1 LF calculated in the 
magnitude range of 20 $\leq$ V $\leq$ 22 is 
$\log \Phi(V) \propto (0.58 \pm0.04)V$, 
which is in good agreement with that expected from a constant 
or slightly declining ($\tau > 8$ Gyr) SFR. 
In comparison, the bar LF is shallower, 
with $\log \Phi(V) \propto (0.42 \pm0.03)V$, indicating 
that its SFR has increased with time.  The SFHs are quantified below.
The LFs clearly suggest the mean age of stars in the bar are younger 
than those in the disk.

A closer look at the red clump 
morphology of the two CMDs is shown in Figure \ref{cmds}c--d.
Although the red clumps have very similar colors, the Disk 1 red clump
is $\approx 0.2$ mag fainter in I than the bar red clump.  This suggests
that the typical red clump star in the disk is older than the 
typical red clump star in the bar (e.g., Cole 1998, Girardi \& Salaris 2001).
This conclusion is supported by the extension of the Disk 1 red clump, but
not the bar red clump,
to fainter and bluer colors:  the classical red horizontal branch 
characteristic of an ancient population. 
In contrast, the bar clump shows a vertical feature at its
blue edge, extending from  $17.4 \leq$  I $\leq 18.8$: the signature
of a stellar population component aged $\approx$ 1 Gyr for 
metallicity Z $\approx$ 0.008 \citep{gir99}, as appropriate for
the metallicity of the young population in the LMC 
([Fe/H] $\approx -0.3$; Hill et al.~1995). 

\begin{figure}
\plotone{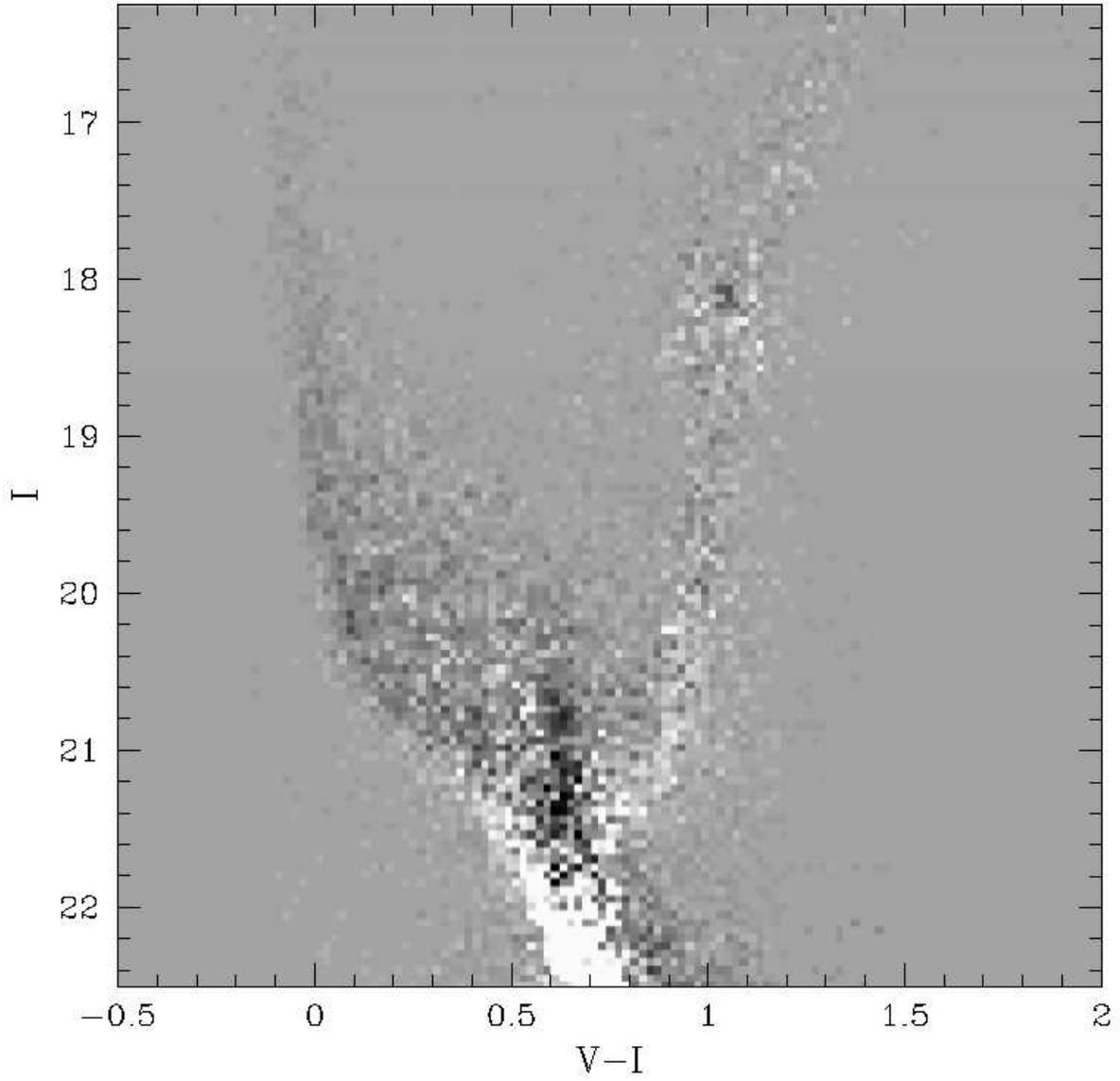}
\caption{ The differential Hess diagram.
Dark regions show areas of higher density in the bar,
white regions show higher density in the disk, and gray indicates
similar densities. The excess population in the bar with ages 
$\approx$ 4 to 6 Gyr is highlighted. \label{difhess}}
\end{figure}

Strong evidence for a younger, burst-dominated
bar can be seen in Figure ~\ref{difhess}, which shows the differential
Hess diagram created by subtracting the scaled disk Hess diagram from
the bar Hess diagram.   The CMDs were binned in 0.02 mag intervals
in V$-$I and 0.05 mag in I, and normalized such that the total number of
stars were equal.
Each pixel in the resulting image has been divided by its Poisson-noise error
to produce Figure~\ref{difhess}.  The blackest pixels represent 
an excess near $3\sigma$ in the bar, and the whitest pixels show
an excess near $5\sigma$ in the disk.
The differences between the fields are dramatically highlighted.
The differential Hess diagram emphasizes two other features that
reinforce the inferences from Figures \ref{cmds}
and \ref{lffig}.
Disk 1 contains more faint subgiants in the range of $20.8 \leq$ I $\leq 21.6$
and $0.7 \leq$ V$-$I $\leq 1.0$, and has a trememendous excess on the
blue side of the main sequence below I $\approx$ 21.4.  Differences in
incompleteness can account for only half of this difference. 
These differences indicate that the disk has a higher fraction of 
older and/or more metal-poor stars than the bar.

\begin{figure}
\plotone{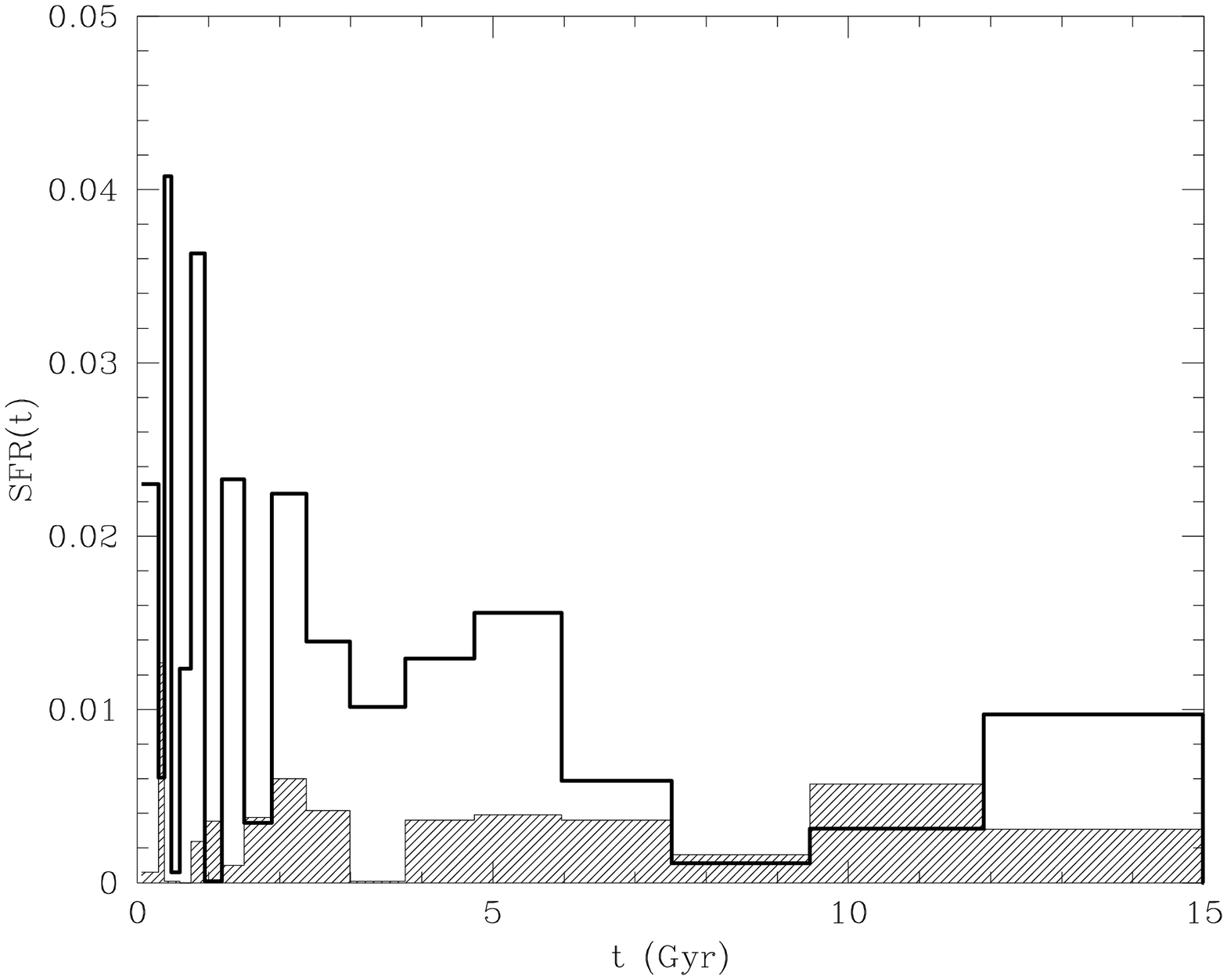}
\caption{ SFHs derived from the main-sequence LFs.
Thick line: SFH bar; shaded line: 10 $\times$ SFH Disk 1.
Units are M$_{\sun}$ yr$^{-1}$ deg$^{-2}$; the
errors in each age bin are $\approx$ $\pm$15\%. \label{sfhfig}}
\end{figure}

Using the adopted distance, reddening, IMF, stellar models, 
and age-metallicity relationship, we have quantify the Disk 1 and Bar 
SFHs by modeling the LF; we do so by minimizing the $\chi^2$ of model 
LFs created by 
sampling the isochrones with a Monte Carlo procedure.  We calculate
the SFRs in logarithmically-spaced time bins and show the results
in Figure~\ref{sfhfig}.  The dominant starbirth event in the bar took
place from 4 to 6 Gyr ago assuming the metallicity increased from
0.004 $\leq$ Z $\leq$ 0.008 during this time.
The metallicity evolution of the bar is poorly known, and it is
the main factor limiting our ability to constrain the exact time and duration 
of the inferred burst of star formation. (We plan to obtain
spectra of individual red giant stars in the bar with the multi-object
spectrograph on the Gemini South 8-meter telescope once it comes online.)
The $\sim$5 Gyr event also generates the excess on the red clump
seen at I = 18.1, V$-$I = 1.05 in Figure~\ref{difhess}.  
During this 2 Gyr window, the average bar SFR was at least twice the
previous average; the bar also experienced
enhanced star-formation $\approx$ 1 to 2 Gyr ago.  The SFR
has also been high for much of the past Gyr, although saturation
limits our ability to constrain it in the last 200 Myr.
The signature morphology of the red clump implies that the $\sim$ 1
Gyr event really was short in duration; in a future paper, we will compare
these two bar fields to two others to explore the possibility that
a dissolved star cluster might give rise to this recent spike in the SFR.   
We estimate from inspection of the isochrones
that the specific ages of the SFH features are subject to a 
10 to 15\% uncertainty, due mainly to the uncertainties in the
adopted age-metallicity relationship.

The derived SFH for the bar predicts that, by mass, 35\% of the stars
are younger than 3 Gyr, and 71\% are younger than 7.5 Gyr. 
For Disk 1, the corresponding fractions at the same ages are
only 19\% and 41\%, respectively.  During the interval 
7.5 Gyr $\leq t \simlt 15$ Gyr, the bar formed 29\% of its stellar
mass, while Disk 1 formed 59\%. The total stellar surface density 
in the bar fields is $\approx 5$ times that of Disk 1.  However, 
if we consider only stars with ages $\geq 7.5$ Gyr, the ratio of the 
surface densities drops to 2.7;  the main uncertainty in the estimate
being the assumed age-metallicity relationship.  This ratio roughly 
agrees with the factor of 2.0 we would expect given that the exponential 
scale length of the disk (excluding the bar) is 2$\fdg$5 
\citep{wn01}. Therefore, the bar and Disk 1 fields
may have shared similar SFHs for $t \geq 7.5$ Gyr.
Disk 1 is found to have experienced a smoother SFH than the
bar, forming a larger fraction of its stars during the first
half of its lifetime.  After an initial epoch of star-formation,
Disk 1 shows only mild variations (a factor $\simlt 2$) in SFR 
until $\approx 3$ Gyr ago.
The SFR seems to have been mildly elevated at t $\approx$ 2 Gyr,
at the same time as the younger enhancement in the bar. 
In contrast to the bar, Disk 1 shows a very low recent SFR.  This
picture of the LMC's SFH will be further sharpened 
through detailed modeling of the CMDs and metallicity
distribution functions in our future papers. 

\section{Discussion}

Our data represent the largest sample of LMC stars
reaching down to the oldest MSTO and beyond.  This has allowed
us to compare the SFHs of the LMC disk and bar 
with unprecedented power.  From analysis of the LFs, 
we find that the disk and bar experienced similar SFHs at
older ages (7.5 to $\sim$ 15 Gyr). However, there is a dramatic
difference in the derived SFH for younger ages. The bar's SFH was dominated
by two distinct enhancements in the SFR from 4 to 6 and 1 to 2 Gyr ago,
while the Disk maintained a nearly constant SFR.
We associate the 4 to 6 Gyr enhancement in the SFR seen {\it only} in the
bar field with the epoch of the formation of the LMC bar.
The timing of the enhancements in the SFR of the bar is intriguing,
because of the similarity they bear to the predicted times
of close interaction between the LMC, SMC, and Milky Way (see, e.g.,
Gardiner et al.~1994).  We hypothesize that tidal interactions
during perigalactic passage $\sim$ 4.5 Gyr ago triggered an instability
in the LMC's gaseous disk that resulted in the formation of the bar.
Large gas flows into the bar during
this interaction could have resulted in the birth of a large
number of stars on bar-type orbits.

How does our derived SFH of the bar compare to previous
estimates based on other WFPC2 data?
\citet{els97} studied a single field near the bar-disk interface and
suggested that the bar formed much later than the disk,
$\approx$ 1 Gyr ago.  Because they were limited to one WFPC2 field,
they based their conclusions on assigning substructures within
the CMD to the bar or disk based on similarity to outer disk fields.
The direct differential comparison provided in our Figure~\ref{difhess}
should be more robust than correlating spatial and CMD structures
within a single field.
\citet{ard97} analyzed a central bar field and found a large fraction of
old stars, with a hiatus in star-formation from 0.5 to 2 Gyr ago.
However, they sampled only $\approx$ 5\% of the number of stars
in our bar field; the resulting uncertainties due to small number
statistics likely account for the discrepancy between their 
conclusion and ours.   \citet{ols99} studied several bar fields
and found the bar to be dominated by stars older than 4 Gyr.  With
our larger sample size, we are able to identify the contribution
of the starbirth event at $\approx$ 5 Gyr to this bar population.
This event seems to be recognizable, at lower signal-to-noise,
within Olsen's Figures 13--14,
although our data suggest a smoother SFH
prior to the 5 Gyr event than was derived by Olsen.  Holtzman et al.~1999
analyzed a single WFPC2 field in the immediate vicinity of our bar
field, and found similar results to \citet{ols99}.  
As we do here, H99 used a differential Hess diagram
to compare the bar to the disk.  The higher fraction of ancient
stars found by H99 may be due in part to the smaller number of
stars analyzed, or to differences in the analysis. We will
investigate this further in future papers.

\acknowledgments

Financial support was provided in part by the NSF through grant
AST-9619460 to TSH, and by NASA through grant GO\#7382 to TSH and JG
from the Space Telescope Science Institute, which is operated by the
Association of Universities for Research in Astronomy, Inc.~under NASA
contract NAS-26555.  This research has made use of NASA's
Astrophysics Data System Abstract Service, the Canadian Astronomical
Data Center,  and the SIMBAD database of the 
Centre de Donn\'{e}es Astronomiques de Strasbourg.

\end{document}